\documentstyle{llncs}

\title{A package TESTAS for checking some kinds of testability}

\author{A.N. Trahtman}
\date{}
\institute{Bar-Ilan University, Dep. of Math. and St., 52900,Ramat Gan,Israel email:trakht@macs.biu.ac.il}
\begin{document}

\maketitle

\centerline{Lecture Notes in Computer Scence 2608(2003), 228-232}

\begin{abstract}
We implement a set of procedures for deciding whether or not a language
given by its minimal automaton or by its syntactic semigroup is locally
testable, right or left locally testable, threshold locally testable,
strictly locally testable, or piecewise
testable.
The  bounds on order of local testability of transition graph
and order of local testability of transition semigroup are also found.
For given $k$, the $k$-testability of transition graph is verified.
Some new effective polynomial time algorithms are used.
These algorithms have been implemented  as  a $C/C ^{++}$ package.
  \end{abstract}

\section*{Introduction}
Locally testable and piecewise testable languages with generalizations
are the best known subclasses of star-free languages with wide spectrum
of applications.

  Membership of a long text in a locally testable language just depends on a
scan of short subpatterns of the text.
  It is best understood in terms of a kind of computational procedure used
 to classify a two-dimensional image: a window of relatively small size
 is moved around on the image and a record is made of the various
 attributes of the image that are detected by what is observed through the
 window. No record is kept of the order in which the attributes are observed,
 where each attribute occurs, or how many times it occurs. We say that a class
 of images is locally testable  if a decision about whether a given image
 belongs to the class can be made simply on the basis of the set of attributes
 that occur.

Kim, McNaughton and McCloskey have found
   necessary and sufficient conditions of local testability
   for the state transition graph $\Gamma$ of
deterministic finite automaton \cite {K91}. By considering the
cartesian product $\Gamma \times \Gamma$,
 we modify these necessary and sufficient conditions and
the algorithms used in the package are based on this approach.

    The locally threshold testable languages were introduced by Beauquier and
 Pin \cite {BP}. These languages generalize the concept of
 locally testable language and have been studied extensively in recent years.

Right [left] local testability was introduced and studied by
K\"{o}nig \cite{Ko} and by Garcia and Ruiz {\cite {GR}}. These
papers use different definitions of the conception and we follow
here \cite{GR}:

 A finite semigroup $S$ is right [left] locally testable iff it is locally
idempotent and locally satisfies the identity $xyx=xy$ [$xyx=yx$].

    We introduced polynomial time algorithms for the right
 [left] local testability problem for transition graph and transition
  semigroup of the deterministic finite automaton.
Polynomial time algorithm verifies transition graph of automaton
with locally idempotent transition semigroup.

There are several systems for manipulating automata and semigroups.
 The list of these systems is following  \cite {FP}
 and preprint of \cite {Ca}:

 REGPACK \cite {Le}
 AUTOMATE \cite {Ch}
 AMoRE \cite {MM}
 Grail \cite {RW}
 The FIRE Engine \cite {Wa}
 LANGAGE \cite {Ca}.
  APL package  \cite {CP}.
 Froidure and Pin package  \cite {FP}.
 Sutner package  \cite {Su}.
 Whale Calf \cite{Ok}.

Some algorithms concerning distinct kinds of testability of finite
automata were implemented by Caron \cite {Ca}, \cite {C2}. His
programs verify piecewise testable, locally testable, strictly and
strongly locally testable languages.

In our package TESTAS (testability of automata and semigroups),
 the area of implemented algorithms
 was essentially extended. We consider important and highly complicated
case of locally threshold testable languages  \cite {TC}.
 The transition semigroups
of  automata are studied in our package at the first time  \cite {Ts}.
Some algorithms (polynomial and even in some way non-polynomial)
check the order of  local testability \cite {Tp}.
We implement a new efficient algorithm for piecewise testability
 improving the time complexity from  $O(n^5)$ \cite {Ca}, \cite {St}
 to $O(n^2)$ \cite {TC}. We consider algorithms for right local testability
($O(n^2)$ time and space complexity), for left local testability
($O(n^3)$ time and space complexity) and the corresponding algorithms
for transition semigroups ($O(n^2)$ time and space complexity).
 The graphs of automata
with locally idempotent transition semigroup are checked too
($O(n^3)$ time complexity). All algorithms dealing with transition semigroup
of automaton have $O(n^2)$ space complexity.

 \section*{Algorithms used in the package}
Let the integer $a$ denote the size of alphabet and let $g$
be the number of nodes. By $n$
let us denote here the size of the semigroup.

 The syntactic characterization of locally threshold
 testable languages was given by Beauquier and Pin \cite {BP}.
From their result follow  necessary and sufficient conditions of
 local threshold testability for transition graph  of DFA
 \cite{TC} and used in our package a polynomial
 time algorithm for the local threshold
testability problem for transition graph and for transition
semigroup of the language.

Let us notice here that the algorithm for transition graph from
\cite{TC} (\cite{Tm}) is valid only for complete graph. Of course,
the general case can be reduced to the case of complete graph by
adding of a sink state. Let us notice also another error from
\cite{TC} (\cite{Tm}): in the Theorem
 16 (17) in the list of the conditions of local threshold testability, the
 property that any $T_{SCC}$ is well defined is missed.

The time complexity of the graph algorithm for local threshold testability
is $O(ag^5)$. The algorithm is based on consideration of the graphs
$\Gamma^2$ and $\Gamma^3$ and therefore has $O(ag^3)$
 space complexity. The time complexity of the semigroup
algorithm is $O(n^3)$.

Polynomial time algorithms for the local testability problem
 for semigroups \cite {Ts} of order $O(n^2)$ and for graphs \cite {TC}
 of order $O(ag^2)$ are implemented in the package too.
We use in our package
 a polynomial time algorithm of worst case asymptotic cost $O(ag^2)$ for
finding the bounds on order of local testability for a given
transition graph of the automaton \cite {Tp} and
 a polynomial time algorithm of worst case asymptotic cost $O(ag^3)$ for
checking the $2$-testability \cite {Tp}. Checking the
$k$-testability for fixed $k$ is polynomial but growing with $k$.
For checking the $k$-testability \cite {Tp}, we use
 an algorithm of worst case asymptotic cost $O(g^3a^{k-2})$.
The order of the last algorithm is growing with $k$ and  so
we have non-polynomial algorithm for finding the order of
local testability.
The algorithms are based on consideration of the graph
$\Gamma^2$ and have $O(ag^2)$ space complexity.
The $1$-testability is verified by help of algorithm of
 cost $O(a^2g)$.

The situation in semigroups is more favorable than in graphs.
 We implement  in our package
 a polynomial time algorithm of worst case asymptotic cost $O(n^2)$ for
finding the order of local testability for a given semigroup \cite
{Ts}.
 The class of locally testable semigroups coincides with the class of strictly
 locally testable semigroups \cite {Tr}, whence the same algorithm
of cost $O(n^2)$ checks strictly locally testable semigroups.

 Stern \cite {St} modified necessary and sufficient conditions
of piecewise testability of DFA (Simon \cite {Si})
 and described a polynomial time algorithm to verify piecewise testability.

 We use in our package a polynomial time algorithm to verify
 piecewise testability of
deterministic finite automaton of worst case asymptotic cost
 $O(ag^2)$ \cite {TC}.
 In comparison, the complexity of Stern's algorithm
\cite {St} is  $O(ag^5)$. Our algorithm uses $O(ag^2)$ space.
 We implement also an algorithm to verify
piecewise testability of a finite semigroup of cost $O(n^2)$

 \section*{Description of the package TESTAS}
 The package includes programs that analyze:

 1) an automaton of the language presented as oriented labeled
graph;

 2) an automaton of the language presented  by its syntactic semigroup,

and find

 3) the direct product of two semigroups or of two graphs,

 4) the syntactic semigroup of an automaton presented by
 its transition graph.

First two programs are written in $C/C ^{++}$ and can by used in
WINDOWS environment. The input file may be ordinary txt file.
We open source file with transition graph or
transition semigroup of the automaton in the standard way and then
check different properties of automaton from menu bar. Both graph
and semigroup are presented on display by help of rectangular table.

 First two numbers in input graph file are the size of alphabet and
the number of nodes.
Transition graph of the automaton is presented by the matrix:

                      \centerline{ nodes X labels}

 where the nodes are presented  by integers from 0 to n-1.
 i-th row of the matrix is a list of successors of i-th node according the
 label in column. The (i,j) cell contains number of the node
 from the end of the edge with label
 from the j-th column and beginning in i-th node.
There exists opportunity to define the number of nodes, size
of alphabet of edge labels and to change values in the matrix.

The input of semigroup algorithms is
Cayley graph of the semigroup presented by the matrix:

 \centerline{ elements X generators}

 where the elements of the semigroup are presented
by integers from 0 to $n-1$ with semigroup generators in the beginning.
i-th row of the matrix is a list of products of i-th element on all generators.

Set of generators is not necessarily minimal, therefore
the multiplication table of the semigroup  (Cayley table) is acceptable too.
 Comments without numerals may be placed in the input file as well.

  The program checks local testability, local threshold testability and
piecewise testability of syntactic semigroup of the language.
 Strictly locally testable and
 strongly locally testable semigroups are verified as well.
 The level of local testability of syntactic semigroup is also found.
Aperiodicity and associative low can be checked too. There exists
possibility to change values of products in the matrix of the
Cayley graph.

The checking of the algorithms is based in particular on the fact
that the considered objects belong to variety and therefore are
closed under direct product. Two auxiliary programs written in C
that find direct product of  two semigroups and of two graphs
belong to the package. The input of semigroup program consists of
two semigroup presented by their Cayley graph with generators in
the beginning of the element list. The result is presented in the
same form and the set of generators of the result is placed in the
beginning of the list of elements. The number of generators of the
result is $n_1g_2 +n_2g_1 - g_1g_2$ where $n_i$ is the size of the
i-th semigroup and $g_i$ is the number of its generators. The
components of direct product of graphs are considered as graphs
with common alphabet of edge labels. The labels of both graphs are
identified according their order. The number of labels is not
necessary the same for both graphs, but the result alphabet used
only common labels from the beginning of both alphabets. Big size
semigroups and graphs can be obtained by help of these programs.

An important verification tool of the package is the possibility
to study both transition graph and semigroup of an automaton.
The program written in C finds syntactic semigroup from the transition graph
of the automaton.

Maximal size of semigroups we consider on standard PC was about some
 thousands elements. Maximal size of considered graphs was about some
 hundreds nodes. The program used in such case memory on hard disc
 and works some minutes.

 \end{document}